\documentclass[sigconf]{acmart}
\usepackage{bm}
\usepackage{url}
\usepackage{xcolor}
\usepackage{bbm}
\usepackage{graphicx}
\usepackage{makecell}
\usepackage{algorithm}
\usepackage{algorithmicx}
\usepackage[noend]{algpseudocode}
\usepackage{caption}
\usepackage{amsmath}
\usepackage{multirow}
\usepackage{enumitem}
\usepackage{subcaption}
\usepackage{tabularx}
\usepackage{longtable}
\usepackage{supertabular}

\usepackage{xspace}
\usepackage{color} 
\usepackage{layouts}
\usepackage[utf8]{inputenc}
\usepackage{eqparbox}
\usepackage{etoolbox} 
\usepackage{float}
\usepackage{tikz}
\usepackage{pgfplots}
\usepackage{filecontents}
\usetikzlibrary{positioning, fit}
\usetikzlibrary{matrix, decorations.pathreplacing,calligraphy}

\renewcommand\footnotetextcopyrightpermission[1]{} 

\AtBeginDocument{%
  \providecommand\BibTeX{{%
    \normalfont B\kern-0.5em{\scshape i\kern-0.25em b}\kern-0.8em\TeX}}}

\setcopyright{acmcopyright}
\copyrightyear{2025}
\acmYear{2025}
\acmDOI{10.1145/1122445.1122456}

\acmConference[CIDR '26]{}{}{}
\acmBooktitle{SIGMOD '25: ACM SIGMOD/PODS International Conference on Management of Data,
  , 2025, SIGMOD, }
\acmPrice{15.00}
\acmISBN{978-1-4503-XXXX-X/18/06}

\begin{document}

    \title{ARCADE: A Real-Time Data System for Hybrid and Continuous Query Processing across Diverse Data Modalities}


\author{Jingyi Yang*, Songsong Mo*, Jiachen Shi, Zihao Yu, Kunhao Shi, Xuchen Ding, Gao Cong}
\thanks{*Both authors contributed equally to this research}
\affiliation{
\institution{Nanyang Technological University, Singapore}
\country{}
}


\newcommand{\blue}[1]{\textcolor{blue}{#1}}

\newcommand{\myparagraph}[1]{\vspace{0.1\baselineskip}\noindent{\textbf{#1.}}~}
\newcommand{\todo}[1]{\textcolor{orange}{#1}}
\begin{abstract}
The explosive growth of multimodal data—spanning text, image, video, spatial, and relational modalities, coupled with the need for real-time semantic search and retrieval over these data—has outpaced the capabilities of existing multimodal and real-time database systems, which either lack efficient ingestion and continuous query capability, or fall short in supporting expressive hybrid analytics. We introduce ARCADE, a real-time data system that efficiently supports high-throughput ingestion and expressive hybrid and continuous query processing across diverse data types. ARCADE introduces unified disk-based secondary index on LSM-based storage for vector, spatial, and text data modalities, a comprehensive cost-based query optimizer for hybrid queries, and an incremental materialized view framework for efficient continuous queries. Built on open-source RocksDB storage and MySQL query engine, ARCADE outperforms leading real-time multimodal data systems by 7.4× on read-heavy and 1.4× on write-heavy workloads.
\end{abstract}




\maketitle

\vspace{-1em}
\section{Introduction}
Data of diverse modalities, \textit{e.g.}, text, image, video, spatial, sensor data, etc., are continuously being generated at an unprecedented rate in the modern world from social media platforms, urban mobility systems, and the financial sector. This exponential growth of multimodal data is coupled with the increasing demand for semantic search and retrieval over multimodal data, enabled by recent advances in large language models (LLMs) and multimodal large language models (MLLMs) that effectively transform these complex problems into queries over vector embeddings. This imposes substantial challenges on modern database systems, especially concerning real-time data ingestion and efficient execution of complex queries over multimodal data, which are essential for responsive user experiences and actionable insights.

A crucial aspect of real-time multimodal data analytics is the efficient support of two essential query types: hybrid queries and continuous queries. Hybrid queries, a more expressive class that subsumes vector queries, enable joint ranking and filtering across multiple data types—including vector, spatial, textual, and other relational attributes. Such cross-modal queries are particularly vital for real-world applications, where semantic search and retrieval—such as in e-commerce recommendations or social media search—often require integrating vector search with filtering or ranking conditions on non-vector data~\cite{chen2024singlestore, wang2021milvus}. Continuous queries, in contrast, are motivated by the growing demand for real-time monitoring and event-driven analytics. These queries automatically execute at fixed user-defined intervals or upon data updates to process real-time data changes and provide up-to-date results. Since the queries re-executed are frequently hybrid in nature, it is essential for modern systems to efficiently support hybrid and continuous queries simultaneously to enable timely, cross-modal insights.

Existing systems tackling these scenarios broadly fall into two categories: multimodal data systems and real-time data systems. Multimodal data systems include traditional databases with extended multimodal data support, such as PostgreSQL~\cite{postgis,pgvector} and DuckDB~\cite{raasveldt2019duckdb}, and data warehouses, such as Databricks Lakehouse~\cite{armbrust2021lakehouse}, Snowflake~\cite{dageville2016snowflake} and BigQuery~\cite{fernandes2015bigquery}. These systems offer robust support for hybrid queries by integrating various data modalities such as vector embeddings and spatial data into the relational model. However, these systems typically lack the capability for high-throughput data ingestion and do not support continuous queries over real-time data. While some data warehouses, \textit{e.g.}, Napa~\cite{agiwal2021napa} are designed to support continuous queries, they compromise data freshness during query execution, limiting their applicability for real-time analytics.

Real-time data systems, commonly employing log-structured merge-tree~\cite{luo20lsmsurvey}, or LSM-tree, storage architectures, prioritize efficient ingestion and processing of live data streams. Early real-time systems, such as Google Bigtable~\cite{chang2008bigtable} and AsterixDB~\cite{alsubaiee2014asterixdb}, primarily focused on relational data, lacking comprehensive support for multimodal data. Specialized real-time vector databases like Milvus~\cite{wang2021milvus}, although efficient at handling vector similarity searches, often neglect data modalities such as  relational and spatial data, thus failing to adequately support hybrid queries. Recent general-purpose real-time vector databases, such as SingleStore-V~\cite{chen2024singlestore, prout2022cloud}, offer improved support for multimodal data and hybrid queries, but remain limited. For instance, it only supports hash indexes on non-vector attributes, which do not support range filters, and it lacks capabilities for jointly ranking results based on both vector and non-vector attribute similarity such as spatial proximity, highlighting significant gaps in supporting comprehensive multimodal analytics.

Given the above limitations of existing multimodal data systems and real-time data systems, there emerges a clear need for a real-time system capable of efficiently supporting hybrid and continuous query processing across diverse multimodal data. To address this gap, we introduce the ARCADE system, a database system designed to meet the demanding requirements for real-time multimodal data analytics. ARCADE leverages an LSM-based storage engine to enable high-throughput data ingestion capabilities and optimizes for two types of hybrid queries: \textbf{i) Hybrid Search Query}, which filters on a combination of multimodal attributes; \textbf{ii) Hybrid Nearest Neighbor Query}, or Hybrid NN Query, which rank results by combining (multi-)vector similarity with non-vector attribute similarity, such as spatial proximity or textual relevance. Additionally, ARCADE seamlessly supports these hybrid queries in a continuous query environment, enabling real-time insights and responsive analytics. Developing such a system involves proposing solutions to address the following three challenges: 

\noindent \textbf{Challenge \#1: Lack of a unified disk-based secondary index framework for different data modalities within LSM-based storage architecture.} While earlier systems like BigTable~\cite{chang2008bigtable} and AsterixDB ~\cite{ alsubaiee2014storage} proposed disk-based secondary index mechanisms for relational and spatial data, these mechanisms do not extend to vector indexes. Current vector index approaches in LSM-based real-time systems are predominantly memory-based~\cite{wang2021milvus, chen2024singlestore}, resulting in substantial memory consumption and compromised ingestion performance due to synchronization issues between vector index and data writes. 
In our own trials, integrating a global in-memory vector index built with standard libraries like FAISS caused the ingestion throughput to drop by as much as $75\times$, indicating severe performance penalty. While system like SingleStore-V~\cite{chen2024singlestore} adopts a per-segment vector index design, each segment's vector index must still be fully loaded into memory before it can be queried, which incurs significant I/O cost for index loading, and also prevents efficient index access via caching when operating under memory constraints. ARCADE addresses this by proposing an efficient disk-based secondary index framework compatible with different data modalities, enabling unified indexing for vector, spatial and text data under LSM-based architecture.

\noindent\textbf{Challenge \#2: Query optimizers are constrained to a narrow subset of hybrid queries.} Existing systems with cost-based query optimizer~\cite{wang2021milvus, chen2024singlestore, pgvector} typically optimize a limited set of hybrid queries, \textit{e.g.}, hybrid NN queries with ranking based on both vector similarity and non-vector attributes similarity are not supported. Moreover, some query optimization strategies do not fully exploit all available indexes, usually only consider either vector index or non-vector indexes in isolation~\cite{chen2024singlestore, wang2021milvus}. ARCADE resolves this limitation by integrating its unified secondary index framework for multimodal data with a robust cost-based query optimizer, which is built upon the MySQL relational query optimizer, and extended to modalities including vector, spatial and text data. This approach allows for cost-based optimization capable of utilizing an optimal combination of secondary indexes to accelerate hybrid queries. Additionally, ARCADE developed a novel query processing algorithm to efficiently handle hybrid NN queries with joint ranking based on (multi-)vector similarity and non-vector attributes similarity, which is not supported by existing systems' query optimizers.

\noindent\textbf{Challenge \#3: Absence of efficient continuous query processing mechanism.} Current real-time systems often lack optimization for continuous queries, resulting in suboptimal real-time performance and responsiveness. ARCADE leverages the key insight that continuous queries—due to their shared query templates—can benefit from reusing intermediate results across executions. Building on this principle, ARCADE proposes an incremental materialized view-based approach, which improves execution efficiency while maintaining data freshness.

\noindent \textbf{Contributions.} This work makes the following contributions.
\vspace{-2em}
\begin{itemize}
    \item We propose ARCADE, a novel database system designed for real-time multimodal data analytics. ARCADE supports high-throughput data ingestion and effectively supports hybrid and continuous queries across diverse data modalities, including vector, spatial, text, and relational data.
    \item We design and implement an efficient and unified disk-based secondary index framework compatible with various data modalities within the LSM-based architecture.
    \item We introduce a comprehensive cost-based query optimizer for hybrid queries, which builds upon MySQL's relational query optimizer and extends its capability to multimodal data. This framework employs cost-based optimization strategies to accelerate the full range of hybrid search queries and hybrid NN queries by leveraging various secondary indexes. 
    \item We propose an incremental materialized view-based approach for efficient continuous query processing, which maintains data freshness and optimizes real-time query performance.
    \item We implement ARCADE on popular open-source systems, \textit{i.e.}, RocksDB as storage and MySQL as query engine. Experiment on real-world datasets shows that ARCADE outperforms leading real-time multimodal data systems by $1.4\times$ and $7.4\times$ on write and read-heavy workloads, respectively.
\end{itemize}

 

\section{Data and Query Definition}
\label{section:data_def}
\subsection{Data \& Index Types} \label{subsection:data-type}
ARCADE fully supports relational data and extends its data model to natively accommodate a broad spectrum of new data modalities, including vector, spatial, text, and general blob data. The multimodal support enables users to store, index, and query heterogeneous real-world data within a single system.
ARCADE supports the declaration of the new data types and indexes as follows:

\noindent \textbf{Vector data.} Declared as \texttt{JSON} type with an explicit vector dimension \texttt{VECTOR\_DIMENSION}. ARCADE supports inverted file vector index with \texttt{VECTOR\_INDEX\_TYPE} option \texttt{`ivf`} for standard IVF index and \texttt{`pqivf`} for IVF index with product quantization.

\noindent \textbf{Blob data.} Used to store general unstructured binary data (e.g., images, videos, or files), declared as \texttt{BLOB}. Unstructured data are usually stored together with the vector embeddings derived and queried through the vector embeddings.

\noindent \textbf{Spatial data.} Declared as \texttt{GEOMETRY} types. ARCADE supports spatial index with \texttt{SPATIAL\_INDEX\_TYPE} option \texttt{`local`} for per-segment index and \texttt{`hybrid`} for hybrid index with additional global index.

\noindent \textbf{Text data.}  Declared as \texttt{TEXT} type. ARCADE supports secondary inverted index for text data.

\vspace{-1em}
\subsection{Supported Queries} \label{subsection:query-type}
ARCADE introduces four expressive query types—two types of hybrid query and two types of continuous query—to enable advanced multimodal data analytics over real-time data. These query types allow users to combine filtering, aggregation, and nearest neighbor search across multimodal data types.



\noindent\textbf{Type 1: Hybrid Search Query} \\
Hybrid search queries enable selection with multiple predicates across relational and non-relational (vector, spatial, or textual) attributes.
\noindent \textit{Example:} Find semantically relevant tweets given a threshold that mention a specific keyword within a geographic region.
{\small
\begin{verbatim}
SELECT t.content,
    ST_Contains(t.coordinate, @region), 
    L2_Distance(t.embedding, LLM(@query_text)) as d1
FROM tweets t
WHERE d1 < @threshold AND t.content LIKE '%keyword%';
\end{verbatim}
}



\noindent\textbf{Type 2: Hybrid NN Query} \\
Hybrid NN queries rank results based on a combination of similarity measures over multiple modalities, such as embedding distance, spatial proximity and textual relevance. They may also involve filter predicate over relational or non-relational attributes.
\noindent \textit{Example:} Find relevant tweets posted during a specific time range based on a weighted sum of spatial proximity and vector similarity.
{\small
\begin{verbatim}
SELECT t.content,
       ST_Distance(t.coordinate, @location) AS d1,
       VECTOR_L2(t.text_embedding, LLM(@query_text)) AS d2
FROM tweets t
WHERE t.time BETWEEN @start_time AND @end_time
ORDER BY d1 + lambda * d2
LIMIT k;
\end{verbatim}
}

\noindent\textbf{Type 3: Continuous SYNC Query} \\
Continuous SYNC queries execute and return up-to-date results over real-time data at fixed user-defined intervals. The corresponding queries may be either Type 1 or 2 above.
\noindent \textit{Example:} Continuously monitor the number of tweets regarding a topic across
different cities and find top cities at 60-second intervals, \textit{e.g.}, advertising campaign monitoring.
{\small
\begin{verbatim}
SELECT c.id, c.name, COUNT(*) as count FROM tweets t
JOIN City as c ON ST_Contains(t.coordinate, c.geom)
WHERE L2_DISTANCE(t.embedding, LLM(@query_text)) < @threshold
GROUP BY c.id
ORDER BY count DESC
SYNC 60 seconds;
\end{verbatim}
}

\noindent\textbf{Type 4: Continuous ASYNC Query} \\
Continuous ASYNC queries automatically re-execute in response to underlying data changes. The corresponding queries may be either Type 1 or 2 above.
\noindent \textit{Example:} Return the most up-to-date tweets regarding a given topic, \textit{e.g.}, monitoring for investment research.
{\small
\begin{verbatim}
SELECT t.content, L2_DISTANCE(t.embedding, LLM(@query_text)) as d1
FROM tweets t
WHERE @Now - time < 60 min 
ORDER BY d1
LIMIT 10
ASYNC;
\end{verbatim}
}
\vspace{-0.5em}
\section{System Architecture}\label{section:system-architecture}
\begin{figure}[ht]
	\centering
        \includegraphics[width=0.99\linewidth]{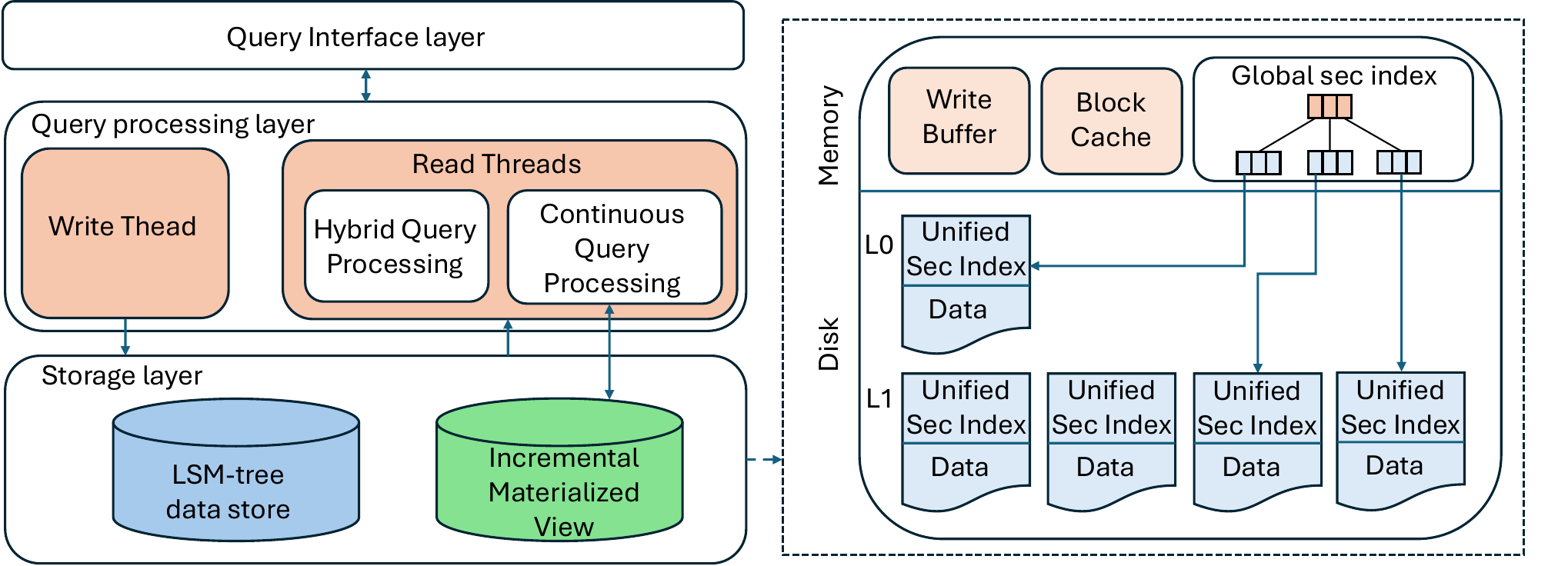}
	\caption{System architecture}
	\label{fig:architecture}
\end{figure}

ARCADE adopts a layered architecture with three layers: query interface, query processing, and storage layer, as shown in Figure~\ref{fig:architecture}.
\vspace{-1.5em}
\vspace{-1.5em}

\noindent \textbf{Query interface layer} serves as the main entry point for user interactions. It receives SQL queries from clients, parses them, and initiates query execution by spawning dedicated read or write threads within the query processing layer.

\noindent \textbf{Query processing layer} is responsible for all query execution tasks. ARCADE separates read and write operations to ensure both high write throughput and low query latency. Given hybrid and continuous queries, ARCADE constructs an optimized execution plan that leverages available secondary indexes and incremental materialized views, efficiently fetching relevant data from the LSM-tree store and materialized views in the storage layer.

\noindent \textbf{Storage layer} persists the data on disk with a partitioned LSM-tree based storage. The LSM-tree data store contains an memory component and a disk component. The memory component includes the skip-list-based write buffer, block cache, and global secondary indexes. The disk component consists of a hierarchy of SST files, which store the unified secondary index and the data.

\vspace{-0.5em}
\section{Secondary Index \& Vector Index} 
\noindent \textbf{Unified Disk-based Secondary Index.}
ARCADE implements an efficient and unified disk-based secondary index framework for multimodal data. Unlike existing LSM-based secondary index for relational data, which maintains the secondary-to-primary key mappings in a separate LSM-tree~\cite{alsubaiee2014asterixdb, chang2008bigtable}, ARCADE integrates the secondary index structure within the primary LSM table to avoid the index navigation cost, which was proposed in NEXT~\cite{next}, but previously limited to a single secondary index on either one-dimensional or multi-dimensional numerical values. In particular, ARCADE adopts a two-level secondary index framework which consists of per-segment components residing in SST files and a global index component in RAM. Each per-segment component stores sorted mappings from secondary attribute values to data block handles, created during SST construction and remaining immutable. The global index, organized as a multi-level tree, maps secondary value ranges to SST index blocks. This design enables efficient SST file pruning and direct query routing during secondary index lookups.

\begin{figure}[ht]
    \includegraphics[trim=0 10 0 15, clip, width=0.9\linewidth]{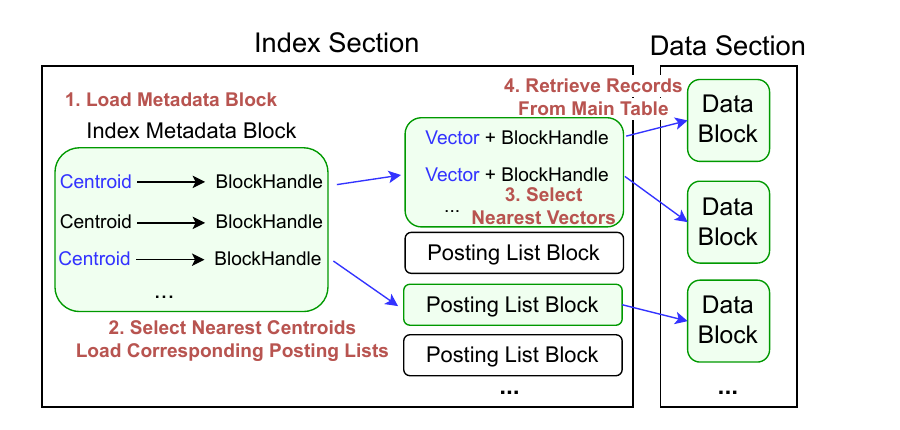}
    \caption{Vector Index Structure}
    \label{fig:vector}
\end{figure}

\noindent \textbf{Vector and Text IVF Index.}
To support a full range of multimodal data, we extend the unified index framework to vector IVF index and text IVF index. The key distinction between our vector IVF index and existing per-segment vector index,\textit{e.g.}, SingleStore-V~\cite{chen2024singlestore}, is that while the existing approach requires each per-segment index to be loaded entirely into memory, our approach enables block-level access,  allowing only a small portion of index blocks to be read per query and facilitating cache reuse across queries. 
As illustrated in Figure~\ref{fig:vector}, the Vector IVF index adopts a two-level approach, the first level consists of index metadata blocks that store IVF centroid-to-posting-list mappings. The second level consists of posting list blocks storing (vector, block handle) pairs. Both types of blocks are implemented as logical SST blocks, the basic write and read unit in RocksDB with cached reads~\cite{rocksdb}. The vector IVF index is built in the background along with SST file construction during memtable flushing or compaction, thus not impacting data ingestion performance as global in-memory vector index approaches do. This design also naturally supports quantization techniques.

\vspace{-1.3em}
\vspace{-1.5em}

Vector index reads leverage a hierarchical iterator structure where per-SST iterators traverse vectors within each SST component, coordinated by a top-level merging iterator using a priority queue. Query execution within each segment follows a three-step process (as shown in Figure~\ref{fig:vector}): first loading centroid metadata, then selectively accessing posting lists for  $n\_probe$ nearest centroids, and finally retrieving records via block handles. This access pattern at finer granularity effectively reduces the memory footprint while improving query performance. Text IVF index is implemented in a similar manner by replacing centroids with the corpus terms. 

\section{Hybrid Query Processing}
\noindent \textbf{Hybrid Search Query Processing.} A major limitation of existing systems is that their query optimizers are unable to generate plans that jointly leverage multiple secondary indexes for hybrid search queries~\cite{wang2021milvus, chen2024singlestore}. Typically, the optimizer only chooses between different plans of leveraging a single index, \textit{e.g.}, pre-filtered approach~\cite{wang2021milvus} uses a non-vector index, while post-filtered approach~\cite{wang2021milvus, chen2024singlestore} uses a vector index. This “index isolation” approach frequently leads to substantial inefficiency, especially for hybrid queries with highly selective filters across multiple modalities.

ARCADE addresses this shortcoming through a comprehensive, cost-based query optimizer integrated with the unified secondary index framework. The optimizer maintains detailed statistics and selectivity estimates for all secondary indexes (vector, spatial, text) in a unified catalog. For each hybrid search query, ARCADE’s optimizer evaluates all viable index access plans using a robust cost model that accounts for index access cost within the LSM layout, expected candidate set size, and residual predicate evaluation overhead. This enables the system to dynamically select the optimal combination of index access paths, \textit{e.g.}, intersecting the bitmap returned from both the spatial and the vector index for queries with relatively low-selectivity filters on both attributes. 
\vspace{-1em}
\begin{algorithm}[H]\small
\caption{Hybrid NN Query Processing}
\label{alg:hybrid-nra}
\begin{algorithmic}[1]
  \Require Sorted index iterators $\mathcal{I}_1,\dots,\mathcal{I}_\ell$;\,
          weights $\lambda_1,\dots,\lambda_\ell$;\,
          $k$
  \Ensure The $k$ objects with the lowest score
          $s(o)=\sum_{j=1}^{\ell}\lambda_j\,d_j(o)$

  \State $R\gets\varnothing$ \Comment{min-heap keyed by {\it upper} bound}
  \State $\textit{LB},\textit{UB}\gets$ empty maps

  \While{\textbf{not} \Call{Stop}{}}
    \For{$j\gets 1$ \textbf{to} $\ell$}
      \State $(o,v)\gets\mathcal{I}_j.$\Call{Next()}{}
      \If{$o\notin\textit{LB}$}
        \State $\textit{LB}[o]\gets 0\;;\; \textit{UB}[o]\gets\infty$
      \EndIf
      \State store $v$ as attribute $j$ of $o$
      \State \Call{UpdateBounds}{$o$}
      \State $R.$\Call{Push\_heap}{$o, \textit{UB}[o]$}
    \EndFor
  \EndWhile

  \State \Return the $k$ objects in $R$ sorted by $\textit{LB}$


  \Function{Stop}{}
    \State \Return $|R|\ge k$
      \textbf{and}
      $\displaystyle\max_{o\in R}\textit{UB}[o]
       \le\min_{o'\notin R}\textit{LB}[o']$
  \EndFunction
\end{algorithmic}
\end{algorithm}
\vspace{-1em}

\noindent \textbf{Hybrid NN Query Processing.} Existing systems typically do not optimize for hybrid NN queries with ranking based on both vector similarity and non-vector attributes similarity, often resorting to costly full scans. To efficiently process hybrid NN queries over multiple modalities, ARCADE adopts an NRA~\cite{fagin2001optimal}-style aggregation algorithm shown in Algorithm~\ref{alg:hybrid-nra}. A key enabler is the unified secondary index framework, which exposes a standardized \textit{Next()} interface for all supported modalities—whether vector IVF index, spatial index, or text index—allowing each index iterator to yield its next smallest distance item (and corresponding block handle) in sorted order. 
This design allows the algorithm to aggregate scores across modalities and efficiently identify the top-k results, simultaneously leveraging all relevant indexes and avoiding full scans.

\section{Continuous Query Processing}
Continuous query optimization often relies on storing and updating query results to improve execution efficiency, typically achieved through indexing continuous queries~\cite{DBLP:journals/tsas/ChenCA23}. However, this traditional approach faces two key challenges: scalability and reusability. As the number of continuous queries issued increases, the storage overhead required to maintain query results grows significantly. Additionally, directly storing complete query results reduces efficiency, as these results can only be reused for specific queries, leading to redundant computations.
\vspace{-1.3em}
\begin{figure}[ht]
	\centering
        \includegraphics[width=0.9\linewidth]{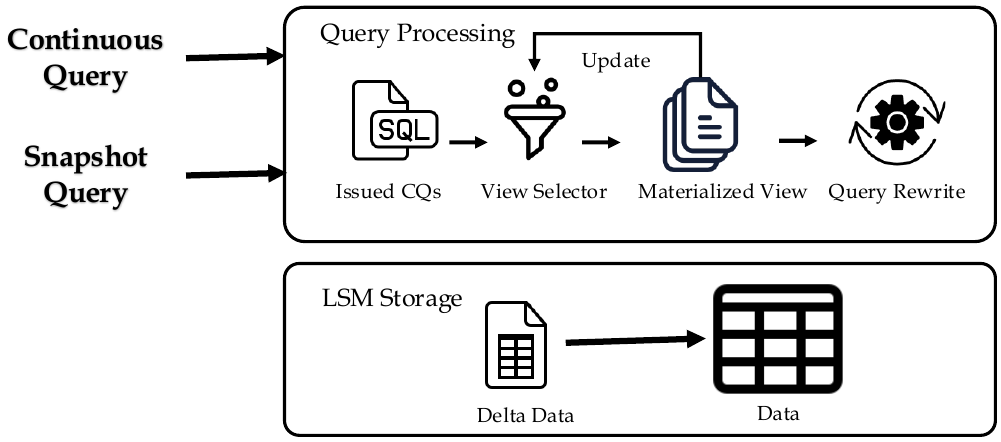}
	\caption{System overview}
	\label{fig:system}
\end{figure}
\vspace{-1.3em}


To address these challenges, we propose an incremental materialized view-based framework that jointly optimizes continuous and snapshot queries. Views are selected from registered continuous queries using a knapsack-based strategy that balances reuse benefit and storage overhead.
The system maintains views incrementally using delta data as new records arrive, avoiding full recomputation. To handle heterogeneous query predicates, we design a unified indexing layer that supports relational, spatial, and vector similarity filters, extending beyond traditional indexing methods which target only a single attribute type~\cite{DBLP:conf/sigmod/SadoghiJ11,DBLP:journals/tsas/ChenCA23}.
Query execution strategies differ by query type: continuous queries reuse linked views via static rewriting at registration time, while snapshot queries select views dynamically at runtime using a greedy heuristic. As shown in Figure~\ref{fig:system}, our framework seamlessly integrates view selection, incremental maintenance, and hybrid query execution, offering both scalability and reusability across diverse query workloads.

\begin{figure*}[!ht]\vspace{-2em}
\begin{subfigure}{.65\textwidth}
\centering
\includegraphics[trim=0 100 0 100, clip, width=\linewidth]{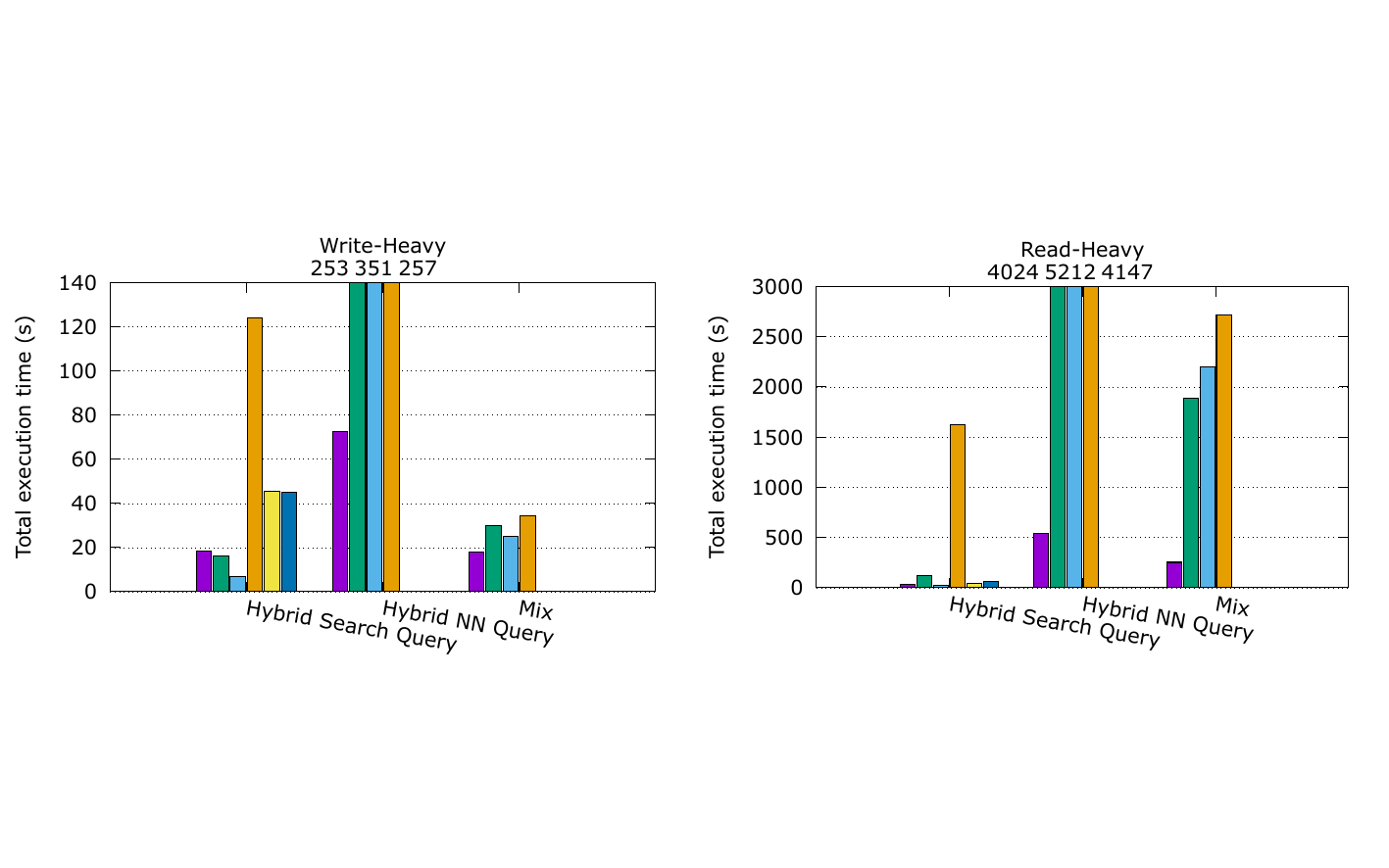}
\end{subfigure}
\begin{subfigure}{.18\textwidth}
    \centering
    \vspace{-3ex}
    \includegraphics[trim=190 90 0 30, clip, width=\linewidth]{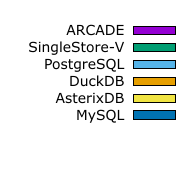}
\end{subfigure}
\caption{Dynamic workload Results}
\label{fig:Dynamic workload Results}\vspace{-2em}
\end{figure*}

In the following sections, we describe the process of materialized view selection and how the selected views are leveraged to improve the execution of both query types.

\myparagraph{Materialized View Selection}
We select and create two types of materialized views:
(1) Spatial Range Views. We cluster similar spatial queries and create one materialized view per cluster. Each view covers a representative spatial region, enabling shared use across queries in the same cluster and reducing redundancy.
(2) Vector NN Views. For vector similarity queries, we cluster queries with similar semantics and create a view for each cluster. Each view materializes top-$xk$ candidates sorted by spatial or vector similarity, and queries reuse these views by re-ranking at runtime to approximate top-$k$ results.

\myparagraph{Incremental Update}
Once views are selected, we maintain them incrementally by applying updates based on incoming delta data, avoiding full recomputation.
To identify affected views, we check which selection predicates of views cover the updated point. For spatial and vector filters, each view defines a coverage region (e.g., hypersphere), stored in an in-memory spatial index (e.g., kd-tree). Upon data updates, we query this index to locate and update all relevant views efficiently.

\myparagraph{Query Execution Using Views}
Query execution involves two steps: view matching and query rewriting.
For continuous queries, views are matched and rewritten at registration time based on shared subqueries, and reused throughout execution. For snapshot queries, matching is done at runtime by analyzing query predicates and views are selected using rule-based heuristics, such as region containment for spatial filters or embedding similarity for vector queries. Rewriting replaces relevant query parts with matched views. Continuous queries use static rewrites, while snapshot queries are dynamically rewritten for each execution.

\vspace{-1.5em}
\section{Experiments} \label{section:exp}
\subsection{TRACY Benchmark} We run the benchmark workloads using our self-built TRACY: \underline{\textbf{T}}weet hyb\underline{\textbf{R}}id \underline{\textbf{A}}nd \underline{\textbf{C}}ontinuous quer\underline{\textbf{Y}} benchmark. The TRACY benchmark evaluates the system performance over real-time hybrid and continuous queries based on a dynamic Tweet analytics scenario.

\noindent\textbf{Dataset.} TRACY comprises three tables: Tweet (33M geo-tagged tweets from Twitter), POI (7M Yelp points-of-interest), and City (186K cities from Open Street Map~\cite{openstreetmap}).
We generate 128-dimensional text embeddings from the Tweet content and POI description.

\noindent\textbf{Query workload.} The TRACY workloads consist of a set of 11 parameterized hybrid query templates, scalable by sampling parameters. The templated queries cover a diverse set of hybrid search queries and hybrid NN queries, with varying combinations of filter predicates and ranking conditions over columns with embedding, spatial and text attributes. 

\noindent \textit{Dynamic Workload generation.} We generate dynamic workloads that mimic real-world analytics scenarios for Twitter data. Prior to workload execution, we pre-load a subset of the Tweet table, \textit{i.e.}, 8 million records, and the complete POI and City tables into the system. We consider two workload scenarios. \textbf{Write-heavy scenario}: 1:9 read-to-write ratio, representing environments with frequent data publication and update; \textbf{Read-heavy scenario}: 9:1 read-to-write ratio, typical of analytics-driven applications with high query volumes. Within each scenario, the workload consists of interleaved writes and queries to simulate concurrent updates and analytical tasks. For both the write-heavy and read-heavy settings, we generate three dynamic workloads, hybrid search query workload, hybrid NN query workload and mixed query workload.

\vspace{-0.5em}
\subsection{Experiment settings}

\noindent \textbf{Baseline systems.} We compare ARCADE with both databases with multimodal data support including PostgreSQL(with PostGIS~\cite{postgis} and pgvector~\cite{pgvector}), MySQL~\cite{mysql}, and DuckDB~\cite{raasveldt2019duckdb}, and real-time systems including SingleStore-V~\cite{chen2024singlestore}, and AsterixDB~\cite{alsubaiee2014asterixdb}. We note that MySQL and AsterixDB do not support vector index, therefore we exclude them from experiments involving hybrid NN queries.

\noindent \textbf{Implementations.} We implement ARCADE on popular open-source systems. We implement the storage based on RocksDB 9.8.0 and the query engine based on MySQL 8.0.32.

\noindent \textbf{Parameters.} We conduct experiments on a server machine
24 Intel i7-13700K CPU and 128GB RAM. We set the read buffer/block cache size of all systems to 512MB. 

\vspace{-0.5em}
\subsection{Dynamic Workload Performance} Figure~\ref{fig:Dynamic workload Results} compares ARCADE with the baseline systems across different dynamic workload scenarios. ARCADE outperforms all baselines on hybrid NN queries, answering them $3.5\times$ and $7.4\times$ faster than the best baseline, SingleStore-V, thanks to the NRA-based query processing algorithm and joint exploitation of multiple indexes. For hybrid search queries, ARCADE only underperforms PostgreSQL, and is slightly slower than SingleStore-V under write-heavy workload. PostgreSQL's lead on hybrid search query is attributed to its highly optimized index implementations and the refined cost model. On mixed workloads, ARCADE delivers a $1.4\times$ and $7.4\times$ speed-up over the best baseline (PostgreSQL and SingleStore-V) on write-heavy and ready-heavy scenarios, respectively.
\vspace{-0.5em}
\subsection{Hybrid Query Performance}
Table~\ref{tab:hybrid-latency} shows the average query latency of ARCADE vs. the baseline systems. ARCADE leads on both hybrid search and NN queries. On hybrid search queries, ARCADE achieves an average latency of 0.12s, outperforming the best baseline AsterixDB by $1.53\times$. PostgreSQL, SingleStore-V and MySQL are $2-3\times$ slower than ARCADE, while DuckDB is over $10\times$ slower. The speedup stems from ARCADE’s comprehensive cost-based query optimizer, which is able to leverage multiple secondary indexes in a single plan; baselines either rely on a single index or resort to full scans. For the more demanding hybrid NN workload, ARCADE beats the fastest baseline, SingleStore-V and DuckDB by $6.8\times$. ARCADE’s NRA-based algorithm, combined with unified index iterators, allows early termination once the top-k bound is satisfied, thereby avoiding the exhaustive candidate scans seen in the baselines.
\vspace{-1em}
\begin{table}[ht]\tiny
  \centering
  \begin{tabular}{lrrrrrr}
    \toprule
    \textbf{Query type} & \textbf{ARCADE} & \textbf{SingleStore-V} & \textbf{PostgreSQL} & \textbf{DuckDB} & \textbf{MySQL} & \textbf{AsterixDB}\\
    \midrule
    Hybrid Search      & \textbf{0.12}  &  0.386 & 0.315 & 2.571 & 0.406 & 0.184\\
    Hybrid NN          & \textbf{0.712} &  4.863 & 11.37 & 4.864 & n/a & n/a\\
    \bottomrule
  \end{tabular}
  \caption{Average query latency (seconds): ARCADE vs.\ Baselines}
  \label{tab:hybrid-latency}
\end{table}
\vspace{-1.5em}
\subsection{Continuous Query Performance}
\begin{figure}[!ht]
  \begin{subfigure}{.23\textwidth}
    \centering
    \includegraphics[width=\linewidth]{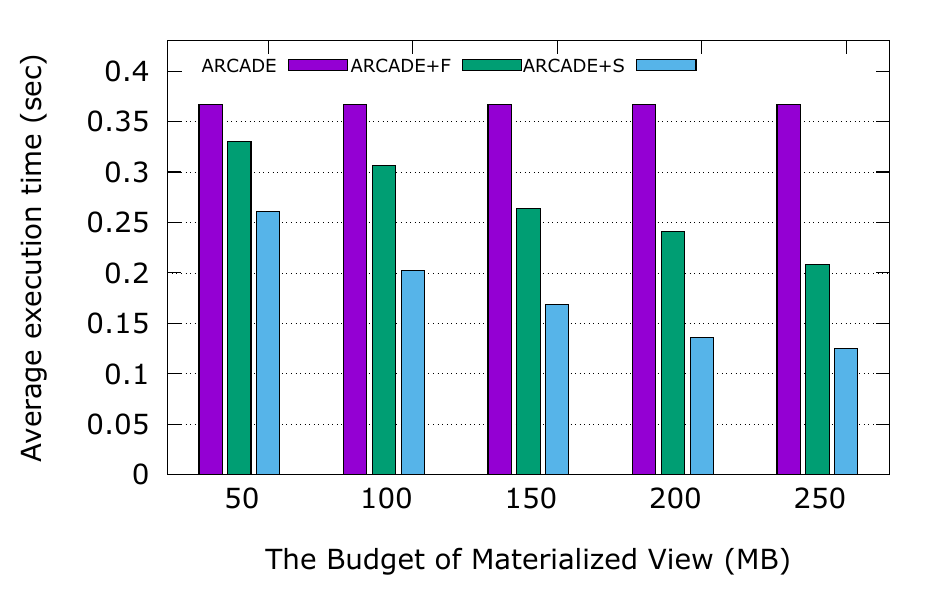}
    \caption{Varying Materialized View Budget}
    \label{fig:cq-budget}
  \end{subfigure}
  \hfill
  \begin{subfigure}{.23\textwidth}
    \centering
    \includegraphics[width=\linewidth]{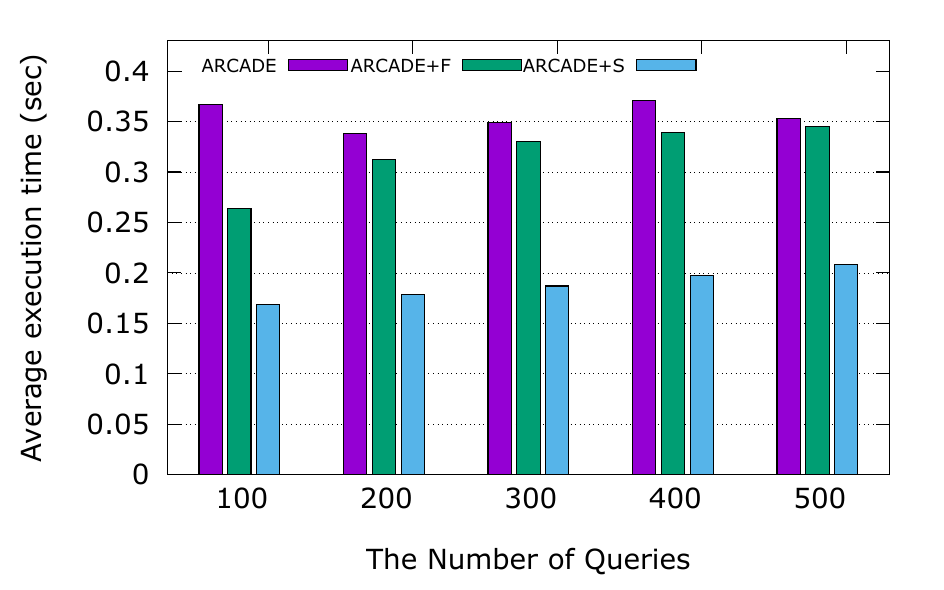}
    \caption{Varying Number of Queries}
    \label{fig:cq-queries}
  \end{subfigure}
  \caption{Continuous query results under different settings.}
  \label{fig:continuous-query}\vspace{-1em}
\end{figure}

We evaluate the performance of our approach under various continuous query workloads, as illustrated in Figure~\ref{fig:continuous-query}. In our experiments, the first baseline (ARCADE) executes queries sequentially without view reuse. We did not compare against external systems, as no existing solution simultaneously supports multi-modal data and both snapshot and continuous queries. Instead, we implemented the core idea from the best available continuous query systems~\cite{DBLP:journals/tsas/ChenCA23} on top of ARCADE, which caches full query results and uses index-based filtering for acceleration. This baseline is denoted as ARCADE+F.
Our proposed method is denoted as ARCADE+S.

Figure~\ref{fig:continuous-query}(a) presents the results for a workload consisting of 100 queries, each producing 1000–2000 result rows. We vary the materialized view memory budget and measure the average execution time. The results demonstrate that our approach consistently achieves lower execution times than both baselines. This improvement is attributed to our materialized view selection strategy, which enables view reuse across multiple queries in the workload, in contrast to prior methods that offer limited coverage.

\vspace{-1.5em}
\vspace{-1em}

In Figure~\ref{fig:continuous-query}(b), we fix the memory budget at 150~MB and vary the number of queries in the workload. While ARCADE+F yields only marginal improvements over the baseline due to its fixed number of maintainable views under the memory constraint, our method provides more substantial performance gains. Notably, as the number of queries increases, the average execution time for ARCADE+F approaches that of the baseline, since the limited number of materialized views maintained can no longer cover most queries. In contrast, our strategy maximizes query coverage, allowing a greater portion of incoming queries to benefit from materialized views.

\vspace{-1em}
\section{RELATED WORK} \label{section:related-work}

\noindent \textbf{Multimodal Data Systems.}
Many traditional database systems has extended support for multimodal data in response to the increasing need for semantic data search and retrieval, including PostgreSQL with PostGIS~\cite{postgis} and pgvector~\cite{pgvector} extensions, MyRocks~\cite{matsunobu2020myrocks} and DuckDB~\cite{raasveldt2019duckdb}. These systems provide robust multimodal query capabilities within the relational model. Cloud data warehouses like Databricks Lakehouse~\cite{armbrust2021lakehouse}, Snowflake~\cite{dageville2016snowflake} and BigQuery~\cite{fernandes2015bigquery} also integrate multiple modalities and support hybrid queries at scale. However, these systems typically target analytical workloads and thus are not optimized for high-throughput data ingestion and lack efficient support for continuous queries over real-time data.

\noindent \textbf{Real-time Data Systems.} Real-time data systems are designed to efficiently ingest and process data streams, Notable early systems such as Google Bigtable~\cite{chang2008bigtable}, HBase~\cite{hbase} and AsterixDB~\cite{alsubaiee2014asterixdb} typically leverage LSM-tree storage architectures~\cite{luo20lsmsurvey} to increase write throughput and focus primarily on relational data, but lacked comprehensive multimodal support. Real-time vector databases, such as Milvus~\cite{wang2021milvus}, specialize in vector similarity search with low-latency, but do not natively support other modalities like relational or spatial data, limiting their ability to handle hybrid queries. While more recent systems like SingleStore-V~\cite{chen2024singlestore, prout2022cloud} incorporate multimodal features and support hybrid queries, their query optimization for hybrid queries remains limited--precluding efficient range filters and joint ranking over attribute similarities. 

\noindent \textbf{Continuous Query System.}
Continuous query systems maintain standing queries over data streams and produce results as data arrives. Early systems, such as Ghanem et al.~\cite{ghanem2008supporting}, proposed SQL-like languages and execution engines, but lack support for analytical queries and reuse mechanisms.
Subsequent work introduced indexing techniques to efficiently match incoming data with continuous queries. Examples include 
BE-tree~\cite{DBLP:conf/sigmod/SadoghiJ11}
for relational streams, and spatial indexes~\cite{DBLP:journals/tsas/ChenCA23} for spatial streams. However, these approaches often assume specific query types (e.g., spatial constraints) and lack unified support for heterogeneous query attributes. In contrast, our framework incrementally maintains a set of shared materialized views to serve both continuous and snapshot queries. It integrates relational, spatial, and vector indexes, enabling effective reuse even when queries vary in type and filtering predicates.
\vspace{-1em}
\section{Conclusion}
In summary, ARCADE is a real-time multimodal data system that unifies support for hybrid and continuous queries. It features a disk-based secondary index framework and a comprehensive cost-based optimizer for hybrid queries. An incremental materialized view mechanism further enhances continuous query performance. Experiments show that ARCADE outperformed existing systems by $7.4\times$ on read-heavy and $1.4\times$ on write-heavy workloads.
\vspace{-1em}

\bibliographystyle{ACM-Reference-Format}
\bibliography{sample-sigconf}


\begin{thebibliography}{24}


\ifx \showCODEN    \undefined \def \showCODEN     #1{\unskip}     \fi
\ifx \showDOI      \undefined \def \showDOI       #1{#1}\fi
\ifx \showISBNx    \undefined \def \showISBNx     #1{\unskip}     \fi
\ifx \showISBNxiii \undefined \def \showISBNxiii  #1{\unskip}     \fi
\ifx \showISSN     \undefined \def \showISSN      #1{\unskip}     \fi
\ifx \showLCCN     \undefined \def \showLCCN      #1{\unskip}     \fi
\ifx \shownote     \undefined \def \shownote      #1{#1}          \fi
\ifx \showarticletitle \undefined \def \showarticletitle #1{#1}   \fi
\ifx \showURL      \undefined \def \showURL       {\relax}        \fi
\providecommand\bibfield[2]{#2}
\providecommand\bibinfo[2]{#2}
\providecommand\natexlab[1]{#1}
\providecommand\showeprint[2][]{arXiv:#2}

\bibitem[ope({[n.\,d.]})]%
        {openstreetmap}
 \bibinfo{year}{[n.\,d.]}\natexlab{}.
\newblock \bibinfo{title}{OpenStreetMap}.
\newblock
\newblock
\newblock
\shownote{\url{https://www.openstreetmap.org}}.


\bibitem[pgv({[n.\,d.]})]%
        {pgvector}
 \bibinfo{year}{[n.\,d.]}\natexlab{}.
\newblock \bibinfo{title}{pgvector}.
\newblock
\newblock
\newblock
\shownote{\url{https://github.com/pgvector/pgvector}}.


\bibitem[pos(2023)]%
        {postgis}
 \bibinfo{year}{2023}\natexlab{}.
\newblock \bibinfo{title}{PostGIS}.
\newblock
\newblock
\newblock
\shownote{\url{https://postgis.net/}}.


\bibitem[mys(2025)]%
        {mysql}
 \bibinfo{year}{2025}\natexlab{}.
\newblock \bibinfo{title}{MySQL}.
\newblock
\newblock
\newblock
\shownote{\url{https://www.mysql.com/}}.


\bibitem[Agiwal et~al\mbox{.}(2021)]%
        {agiwal2021napa}
\bibfield{author}{\bibinfo{person}{Ankur Agiwal}, \bibinfo{person}{Kevin Lai}, {et~al\mbox{.}}} \bibinfo{year}{2021}\natexlab{}.
\newblock \showarticletitle{Napa: Powering scalable data warehousing with robust query performance at Google}.
\newblock \bibinfo{journal}{\emph{PVLDB}} \bibinfo{volume}{14}, \bibinfo{number}{12} (\bibinfo{year}{2021}), \bibinfo{pages}{2986--2997}.
\newblock


\bibitem[Alsubaiee et~al\mbox{.}(2014a)]%
        {alsubaiee2014asterixdb}
\bibfield{author}{\bibinfo{person}{Sattam Alsubaiee}, \bibinfo{person}{Yasser Altowim}, \bibinfo{person}{Hotham Altwaijry}, {et~al\mbox{.}}} \bibinfo{year}{2014}\natexlab{a}.
\newblock \showarticletitle{AsterixDB: {A} Scalable, Open Source {BDMS}}.
\newblock \bibinfo{journal}{\emph{PVLDB}} \bibinfo{volume}{7}, \bibinfo{number}{14} (\bibinfo{year}{2014}), \bibinfo{pages}{1905--1916}.
\newblock


\bibitem[Alsubaiee et~al\mbox{.}(2014b)]%
        {alsubaiee2014storage}
\bibfield{author}{\bibinfo{person}{Sattam Alsubaiee}, \bibinfo{person}{Alexander Behm}, {et~al\mbox{.}}} \bibinfo{year}{2014}\natexlab{b}.
\newblock \showarticletitle{Storage management in AsterixDB}.
\newblock \bibinfo{journal}{\emph{PVLDB}} \bibinfo{volume}{7}, \bibinfo{number}{10} (\bibinfo{year}{2014}), \bibinfo{pages}{841--852}.
\newblock


\bibitem[Apache(2019)]%
        {hbase}
\bibfield{author}{\bibinfo{person}{Apache}.} \bibinfo{year}{2019}\natexlab{}.
\newblock \bibinfo{title}{Apache HBase}.
\newblock
\newblock
\newblock
\shownote{\url{https://hbase.apache.org/}}.


\bibitem[Armbrust et~al\mbox{.}(2021)]%
        {armbrust2021lakehouse}
\bibfield{author}{\bibinfo{person}{Michael Armbrust}, \bibinfo{person}{Ali Ghodsi}, \bibinfo{person}{Reynold Xin}, \bibinfo{person}{Matei Zaharia}, {et~al\mbox{.}}} \bibinfo{year}{2021}\natexlab{}.
\newblock \showarticletitle{Lakehouse: a new generation of open platforms that unify data warehousing and advanced analytics}. In \bibinfo{booktitle}{\emph{Proceedings of CIDR}}, Vol.~\bibinfo{volume}{8}. \bibinfo{pages}{28}.
\newblock


\bibitem[Chang et~al\mbox{.}(2008)]%
        {chang2008bigtable}
\bibfield{author}{\bibinfo{person}{Fay Chang}, \bibinfo{person}{Jeffrey Dean}, {et~al\mbox{.}}} \bibinfo{year}{2008}\natexlab{}.
\newblock \showarticletitle{Bigtable: A distributed storage system for structured data}.
\newblock \bibinfo{journal}{\emph{TOCS}} \bibinfo{volume}{26}, \bibinfo{number}{2} (\bibinfo{year}{2008}), \bibinfo{pages}{1--26}.
\newblock


\bibitem[Chen et~al\mbox{.}(2024)]%
        {chen2024singlestore}
\bibfield{author}{\bibinfo{person}{Cheng Chen}, \bibinfo{person}{Chenzhe Jin}, \bibinfo{person}{Yunan Zhang}, {et~al\mbox{.}}} \bibinfo{year}{2024}\natexlab{}.
\newblock \showarticletitle{Singlestore-v: An integrated vector database system in singlestore}.
\newblock \bibinfo{journal}{\emph{PVLDB}} \bibinfo{volume}{17}, \bibinfo{number}{12} (\bibinfo{year}{2024}), \bibinfo{pages}{3772--3785}.
\newblock


\bibitem[Chen et~al\mbox{.}(2023)]%
        {DBLP:journals/tsas/ChenCA23}
\bibfield{author}{\bibinfo{person}{Zhida Chen}, \bibinfo{person}{Gao Cong}, {and} \bibinfo{person}{Walid~G. Aref}.} \bibinfo{year}{2023}\natexlab{}.
\newblock \showarticletitle{{STAR:} {A} Cache-based Stream Warehouse System for Spatial Data}.
\newblock \bibinfo{journal}{\emph{TSAS}} \bibinfo{volume}{9}, \bibinfo{number}{4} (\bibinfo{year}{2023}), \bibinfo{pages}{28:1--28:27}.
\newblock


\bibitem[Dageville et~al\mbox{.}(2016)]%
        {dageville2016snowflake}
\bibfield{author}{\bibinfo{person}{Benoit Dageville}, \bibinfo{person}{Thierry Cruanes}, \bibinfo{person}{Marcin Zukowski}, \bibinfo{person}{Vadim Antonov}, \bibinfo{person}{Artin Avanes}, {et~al\mbox{.}}} \bibinfo{year}{2016}\natexlab{}.
\newblock \showarticletitle{The snowflake elastic data warehouse}. In \bibinfo{booktitle}{\emph{SIGMOD}}. \bibinfo{pages}{215--226}.
\newblock


\bibitem[Facebook(2023)]%
        {rocksdb}
\bibfield{author}{\bibinfo{person}{Facebook}.} \bibinfo{year}{2023}\natexlab{}.
\newblock \bibinfo{title}{RocksDB}.
\newblock
\newblock
\newblock
\shownote{\url{https://github.com/facebook/rocksdb}}.


\bibitem[Fagin et~al\mbox{.}(2001)]%
        {fagin2001optimal}
\bibfield{author}{\bibinfo{person}{Ronald Fagin}, \bibinfo{person}{Amnon Lotem}, {and} \bibinfo{person}{Moni Naor}.} \bibinfo{year}{2001}\natexlab{}.
\newblock \showarticletitle{Optimal aggregation algorithms for middleware}. In \bibinfo{booktitle}{\emph{PODS}}. \bibinfo{pages}{102--113}.
\newblock


\bibitem[Fernandes and Bernardino(2015)]%
        {fernandes2015bigquery}
\bibfield{author}{\bibinfo{person}{S{\'e}rgio Fernandes} {and} \bibinfo{person}{Jorge Bernardino}.} \bibinfo{year}{2015}\natexlab{}.
\newblock \showarticletitle{What is bigquery?}. In \bibinfo{booktitle}{\emph{IDEAS}}. \bibinfo{pages}{202--203}.
\newblock


\bibitem[Ghanem et~al\mbox{.}(2008)]%
        {ghanem2008supporting}
\bibfield{author}{\bibinfo{person}{Thanaa~M Ghanem} {et~al\mbox{.}}} \bibinfo{year}{2008}\natexlab{}.
\newblock \showarticletitle{Supporting views in data stream management systems}.
\newblock \bibinfo{journal}{\emph{TODS}} \bibinfo{volume}{35}, \bibinfo{number}{1} (\bibinfo{year}{2008}), \bibinfo{pages}{1--47}.
\newblock


\bibitem[Luo and Carey(2020)]%
        {luo20lsmsurvey}
\bibfield{author}{\bibinfo{person}{Chen Luo} {and} \bibinfo{person}{Michael~J. Carey}.} \bibinfo{year}{2020}\natexlab{}.
\newblock \showarticletitle{LSM-based storage techniques: a survey}.
\newblock \bibinfo{journal}{\emph{{VLDB} J.}} \bibinfo{volume}{29}, \bibinfo{number}{1} (\bibinfo{year}{2020}), \bibinfo{pages}{393--418}.
\newblock


\bibitem[Matsunobu et~al\mbox{.}(2020)]%
        {matsunobu2020myrocks}
\bibfield{author}{\bibinfo{person}{Yoshinori Matsunobu}, \bibinfo{person}{Siying Dong}, {and} \bibinfo{person}{Herman Lee}.} \bibinfo{year}{2020}\natexlab{}.
\newblock \showarticletitle{Myrocks: Lsm-tree database storage engine serving facebook's social graph}.
\newblock \bibinfo{journal}{\emph{Proceedings of the VLDB Endowment}} \bibinfo{volume}{13}, \bibinfo{number}{12} (\bibinfo{year}{2020}), \bibinfo{pages}{3217--3230}.
\newblock


\bibitem[Prout et~al\mbox{.}(2022)]%
        {prout2022cloud}
\bibfield{author}{\bibinfo{person}{Adam Prout}, \bibinfo{person}{Szu-Po Wang}, \bibinfo{person}{Joseph Victor}, \bibinfo{person}{Zhou Sun}, \bibinfo{person}{Yongzhu Li}, {et~al\mbox{.}}} \bibinfo{year}{2022}\natexlab{}.
\newblock \showarticletitle{Cloud-native transactions and analytics in singlestore}. In \bibinfo{booktitle}{\emph{SIGMOD}}. \bibinfo{pages}{2340--2352}.
\newblock


\bibitem[Raasveldt and M{\"u}hleisen(2019)]%
        {raasveldt2019duckdb}
\bibfield{author}{\bibinfo{person}{Mark Raasveldt} {and} \bibinfo{person}{Hannes M{\"u}hleisen}.} \bibinfo{year}{2019}\natexlab{}.
\newblock \showarticletitle{Duckdb: an embeddable analytical database}. In \bibinfo{booktitle}{\emph{SIGMOD}}. \bibinfo{pages}{1981--1984}.
\newblock


\bibitem[Sadoghi et~al\mbox{.}(2011)]%
        {DBLP:conf/sigmod/SadoghiJ11}
\bibfield{author}{\bibinfo{person}{Mohammad Sadoghi} {et~al\mbox{.}}} \bibinfo{year}{2011}\natexlab{}.
\newblock \showarticletitle{BE-tree: an index structure to efficiently match boolean expressions over high-dimensional discrete space}. In \bibinfo{booktitle}{\emph{SIGMOD}}. \bibinfo{publisher}{{ACM}}, \bibinfo{pages}{637--648}.
\newblock


\bibitem[Shi et~al\mbox{.}(2025)]%
        {next}
\bibfield{author}{\bibinfo{person}{Jiachen Shi}, \bibinfo{person}{Jingyi Yang}, \bibinfo{person}{Gao Cong}, {and} \bibinfo{person}{Xiaoli Li}.} \bibinfo{year}{2025}\natexlab{}.
\newblock \showarticletitle{NEXT: A New Secondary Index Framework for LSM-based Data Storage}.
\newblock \bibinfo{journal}{\emph{SIGMOD}} \bibinfo{volume}{3}, \bibinfo{number}{3} (\bibinfo{year}{2025}), \bibinfo{pages}{1--25}.
\newblock


\bibitem[Wang et~al\mbox{.}(2021)]%
        {wang2021milvus}
\bibfield{author}{\bibinfo{person}{Jianguo Wang} {et~al\mbox{.}}} \bibinfo{year}{2021}\natexlab{}.
\newblock \showarticletitle{Milvus: A purpose-built vector data management system}. In \bibinfo{booktitle}{\emph{SIGMOD}}. \bibinfo{pages}{2614--2627}.
\newblock


\end{thebibliography}
\end{document}